\documentclass[11pt,a4paper]{article}
\pdfoutput=1
\usepackage{jcappub}

\usepackage[section]{placeins}
\usepackage{bm}
\usepackage{color}
\usepackage{arydshln}

\def\vp{{\varphi}}

\def\cs2{c_{\rm{s}}^2}

\newcommand\calh{\mathcal{H}}
\newcommand\mpl{m_{\rm Pl}}

\renewcommand\({\left(}
\renewcommand\){\right)}

\newcommand\be{\begin{equation}}
\newcommand\ee{\end{equation}}
\newcommand\bea{\begin{eqnarray}}
\newcommand\eea{\end{eqnarray}}
\newcommand\eq[1]{Eq.~(\ref{#1})}
\newcommand\eqs[2]{Eqs.~(\ref{#1}) and (\ref{#2})}

\newcommand\fig[1]{Fig.~ \ref{#1}}

\newcommand\eps{\epsilon}

\newcommand{\bk}{\mathbf{\rm{k}}}

\begin{document}

\title{Observable induced gravitational waves from an early matter phase}
\author[a]{Laila Alabidi}
\emailAdd{laila@yukawa.kyoto-u.ac.jp}
\author[b,c]{Kazunori Kohri}
\emailAdd{kohri@post.kek.jp}
\author[a]{Misao Sasaki}
\emailAdd{misao@yukawa.kyoto-u.ac.jp}
\author[d]{Yuuiti Sendouda}
\emailAdd{sendouda@cc.hirosaki-u.ac.jp}

\affiliation[a]{Yukawa Institute for Theoretical Physics, Kyoto University, Kyoto 606-8502, Japan}
\affiliation[b]{Cosmophysics Group, Theory Center, IPNS, KEK, Tsukuba 305-0801, Japan}
\affiliation[c]{The Graduate University for Advanced Study (Sokendai), Tsukuba 305-0801, Japan}
\affiliation[d]{Graduate School of Science and Technology, Hirosaki University, Hirosaki, Aomori 036-8561, Japan}

\abstract{Assuming that inflation is succeeded by a phase of matter domination, which corresponds
to a low temperature of reheating $T_r<10^9\rm{GeV}$, we evaluate the spectra
of gravitational waves induced in the post-inflationary universe. We work with models of hilltop-inflation with an enhanced
primordial scalar spectrum on small scales, which can potentially lead to the formation of primordial black holes.
We find that a lower reheat temperature leads to the production of gravitational waves with energy densities
within the ranges of both space and earth based gravitational wave detectors.
}
\maketitle

\section{Introduction}

Induced gravitational waves are produced as a result of the interaction between scalar perturbations 
at second order in the post-inflationary universe. The amplitude of their spectra is dependent 
on the square of the primordial scalar spectrum \cite{Matarrese:1993zf,Matarrese:1997ay,Mollerach:2003nq,Ananda:2006af, Baumann:2007zm}, and a 
relatively large induced gravitational wave spectrum is expected from the generation of Primordial Black Holes (PBHs) 
\cite{Saito:2008jc,Saito:2009jt,Bugaev:2009kq,Bugaev:2010bb}. 
In a previous paper Ref.~\cite{Alabidi:2012ex} we evaluated the spectra of induced gravitational waves 
generated during a radiation dominated era from the hilltop-type and running mass models, which have been 
shown to be the only models which can lead to Primordial Black Holes \cite{Kohri:2007gq, Drees:2011yz}. 
We showed that these models lead to an induced gravitational wave signature within the sensitivity ranges of planned  
gravitational wave detectors DECIGO and BBO \cite{Seto:2001qf,decigo,Kudoh:2004he}. 
We also found that the running mass model predicted spectra within the
sensitivity of eLISA \cite{lisa,AmaroSeoane:2012km}under the proviso that inflation is terminated early, 
with the intriguing factor that if we could motivate $N\ll40$ we would
get a detectable signature of PBHs with a mass compatible with Dark Matter. 

In this paper, we assume that the universe undergoes a phase of early matter domination \cite{Seto:2003kc,Boyle:2007zx,Nakayama:2008ip,Nakayama:2008wy}
\footnote{other scenarios are also possible \cite{Joyce:1996cp, Seto:2003kc,Boyle:2007zx, Hidalgo:2011fj}},
which lowers the reheat temperature as well as the number of allowed $e-$folds of inflation.
The source term during the matter phase is constant, and as a result of the absence of pressure, 
the  density contrast grows. This may result in perturbations entering the non-linear regime and 
decoupling from the Hubble flow. In our analysis we only assume a  linear evolution of perturbations, 
and we account for the non-linear evolution by cutting off our analysis at some critical scale.  
To calculate this critical scale we begin by stating that perturbations in the early matter phase evolve as
\be
\frac{\delta\rho_m}{\rho}=\frac{2}{3\mathcal{H}^2}\nabla^2\Phi
\ee
where $\rho_m$ is the energy density of matter, $\calh$ is the conformal Hubble parameter 
and $\Phi$ is the gravitational potential. The evolution of the perturbations is therefore 
linear until the density contrast becomes of order unity which occurs at the scale \cite{Assadullahi:2009jc}:
\be
\label{eq:knl}
k_{NL}\sim \mathcal{P}_\zeta^{-1/4}k_r
\ee
where $k_r$ is the scale which re-enters the horizon at the time of reheating, $\mathcal{P}_\zeta$ 
is the primordial spectrum and $k_{NL}$ is the critical scale at which we terminate our calculation.

This paper is organised as follows, in section \ref{sec:inf} we review 
the parameters of inflation, in section \ref{sec:k_r} 
we present the thermal history of the universe, relating the temperature of reheating to 
the relevant scale of reheating, in section \ref{sec:pbh} we caclulate the bounds on the primordial spectrum from
PBHs, in section \ref{sec:MD} we review the spectrum of 
induced gravitational waves produced during the early matter phase,  in section \ref{sec:models} we review the models of inflation 
that can lead to a detectable limit of induced gravitational waves and penultimately in section \ref{sec:results} 
we present the results with the final discussion presented in section \ref{sec:disc}.

The following conventions are utilised in this paper: $\tau$ refers to conformal
time and is related to proper time $t$ as $d\tau=dt/a$, $a$ is the scale factor, 
and the conformal
Hubble parameter $\calh$ is related to the Hubble parameter $H\equiv\dot{a}/a$ as $\calh=aH$.
Scales are denoted by $k$, are given in units of inverse megaparsec $\rm{Mpc}^{-1}$
and are related to physical frequency $f$ as $f=ck/(2a\pi)$ where $c$ is the speed 
of light. We assume a radiation dominated universe at the time of the formation 
of the gravitational waves, in which case we have $a=a_0(\tau/\tau_0)$, $\calh=\tau^{-1}$,
and the scale at re-entry is $k=\tau^{-1}$. 

\section{Inflationary Parameters}\label{sec:inf}
Models of inflation can be parametrised by the  slow roll parameters \cite{Lyth:2009zz} :
\bea\label{SR}
\eps&=&\frac{\mpl^2}{2}\(\frac{V_{,\vp}}{V}\)^2\nonumber\\
\eta&=&\mpl^2\frac{V_{,\vp\vp}}{V}\nonumber\\
\xi^2&=&\mpl^4\frac{V_{,\vp}V_{,\vp\vp\vp}}{V^2}
\eea
where $V$ is the potential, and
derivatives are with respect to the inflaton field $\vp$. These
are related to the observational parameters,
the spectral index $n_s$, the running of the spectral index $n_s'$ and
the scalar spectrum $\mathcal{P}_\zeta$ as:

\bea\label{para}
n_s&=&1+2\eta-6\eps\nonumber\\
n_s'&=&16\eps\eta-24\eps^2-2\xi^2\nonumber\\
\mathcal{P}_\zeta&=&\frac{1}{24\pi^2\mpl^4}\frac{V}{\eps}
\eea

We use a time re-parametrisation, $N=\ln\(\frac{a_e}{a_*}\)$, where the subscripts $e$ and $*$ denote the end
of inflation and the time of horizon exit respectively. This is related
to the potential in the slow roll limit as:

\be\label{NSR}
N\simeq\mpl^{-2}\int_{\varphi_e}^{\varphi_*}\frac{V}{V'}d\varphi
\ee
and to the scale at horizon exit as \cite{Lyth:2009zz}:
\be\label{Nk}
N(k_0)-N(k)=\ln\(\frac{0.002}{k}\).
\ee
where $k_0=0.002\rm{Mpc}^{-1}$ is the pivot scale, and in this paper we effectively take $N(k_0)=0$.

We use the latest data release from the WMAP mission \cite{Hinshaw:2012fq,Bennett:2012fp}, for the
WMAP data combined with BAO and H0 data with a null tensor prior.
Throughout this paper we take $n_s=0.96$ and $n_s'\leq0.0062$.
\section{The temperature of reheating}
\label{sec:k_r}
The number of $e-$folds can be related to the temperature of reheating $T_r$ as \cite{Liddle:2003as,Dodelson:2003vq}
\be
N=56-\frac{2}{3}\ln\left(\frac{10^9\rm{GeV}}{T_{r}}\right)
\ee
where we have taken the energy scale of inflation to be the SUSY GUT scale $\sim10^{16} \rm{GeV}$. 
Assuming SUSY means that $N_{max}=56$, otherwise we can push this estimate up to $N_{max}\sim60$. 
In this work we are only interested in reducing the number of $e-$folds via the inclusion of an early matter phase. 
The thermal history of the universe can support a reheat temperature down to $1 \rm{MeV}$, as this is the temperature
below which Neutrinos fail to thermalise and affect big-bang nucleosynthesis\cite{Kawasaki:1999na,Kawasaki:2000en,Hannestad:2004px,Ichikawa:2005vw}  
and hence $N\gtrsim37$.
To relate $T_r$ to the scale $k_r$ we assume the conservation of entropy which gives
\bea
H &=& g_{*s}(T_r)^{1/2}  \frac{T_r^2}{ M_G}\nonumber\\
\frac{a}{a_0}&=& \frac{T_0}{T_r}   [g_{*s}(T_0) / g_{*s}(T_r)]^{1/3}\nonumber\\
\eea
where $g_{*s}$ is the number of degrees of freedom, $M_G$ is the gravitational scale ($M_G= M_p / \sqrt{8 \pi} \simeq 2.4 \times10^{18} {\rm GeV}$ )
 and since $k=aH/a_0$ we get the scale which re-enters the horizon at the end of reheating $k_r$:
\bea
k_r&\sim&1.7\times10^{16}\rm{Mpc}^{-1}\left(\frac{T_r}{10^9\rm{GeV}}\right)\left(\frac{g_{*s}}{106.75}\right)^{1/6}
\eea

\section{The spectrum of primordial black holes}
\label{sec:pbh}
If the primordial spectrum of perturbations towards the end of inflation is large enough, 
i.e if the density contrast exceeds $\delta\approx 1/3$, perturbations can collapse to form primordial
black holes. Based on this, constraints can be placed on the spectrum based on astrophysical phenomena \cite{Carr:2009jm,Josan:2009qn}.
In our previous paper, we numerically converted the mass fraction of the PBHs into a power spectrum. 
To perform this analysis we assumed a gaussian distributed energy perturbation \footnote{see Ref.~\cite{Bullock:1996at,Ivanov:1997ia,PinaAvelino:2005rm,Chongchitnan:2006wx,Hidalgo:2007vk,Bugaev:2011wy,Byrnes:2012yx} for
alternative scenarios} and a very large reheat temperature $T_r\gg10^{10}\rm{GeV}$. In this paper, we update this calculation for lower reheat temperatures.

\subsection{$ M_\mathrm{BH}(T) $ and $ k(M_\mathrm{BH}) $}
%
The comoving wavenumber corresponding to the Hubble radius at temperature $ T $ is
\begin{equation}
k
= \frac{a\,H}{c}
\approx
  1.71 \times 10^{16}\,\mathrm{Mpc}^{-1}\,
  \left(\frac{g_{*s}}{106.75}\right)^{1/6}\,
  \left(\frac{T}{10^9\,\mathrm{GeV}/k_\mathrm B}\right)\,.
\end{equation}
On the other hand, the mass of PBHs produced at temperature $ T $ is
\bea
M_\mathrm{BH}
&=& \gamma\,\frac{4 \pi}{3}\,\frac{\rho}{H^3}
= \frac{3\,\sqrt5}{4 \pi^{3/2}\,G^{3/2}}\,\gamma\,g_{*s}^{-1/2}\,T^{-2}\nonumber\\
&\approx &
  0.916\,\times 10^{-20}\,M_\odot\,
  \left(\frac{\gamma}{(1/\sqrt3)^3}\right)\,
  \left(\frac{g_{*s}}{106.75}\right)^{-1/2}\,
  \left(\frac{T}{10^9\,\mathrm{GeV}}\right)^{-2}\,,
\eea
where $ \gamma $ is a numerical factor and $ M_\odot = 1.989 \times 10^{33}\,\mathrm g $\,.
Eliminating $ T $, we find
\begin{equation}
\begin{aligned}
k
&
= \frac{\pi^{3/4}\,g_{*s0}^{1/3}\,T_{\gamma0}}{45^{1/4}\,G^{1/4}}\,
  \frac{g_{*s}^{1/4}}{g_{*s}^{1/3}\,\sqrt{M_\mathrm{BH}/\gamma}} \\
&
\approx
  1.71 \times 10^{16}\,\mathrm{Mpc}^{-1}\,
  \left(\frac{g_{*s}}{106.75}\right)^{-1/12}\,
    \left(\frac{\gamma}{(1/\sqrt3)^3}\right)^{1/2}\,
  \left(\frac{M_\mathrm{BH}}{0.916 \times 10^{-20}\,M_\odot}\right)^{-1/2}\,.
\end{aligned}
\end{equation}

In our numerical calculation we adopt $ \gamma = 1 $ and $g_{*s} = 106.75 $ for whatever values of $ T $\,, which implies
\begin{equation}
M_\mathrm{BH}
\approx
  0.946 \times 10^{28}\,\mathrm g\,
  \left(\frac{T}{10^2\,\mathrm{GeV}}\right)^{-2}\,,
\quad
k
\approx
  1.67 \times 10^9\,\mathrm{Mpc}^{-1}\,
  \left(\frac{M_\mathrm{BH}}{10^{28}\,\mathrm g}\right)^{-1/2}\,.
\end{equation}
These are only precise for $ M_\mathrm{BH} \lesssim 10^{28}\,\mathrm g $ but the error will not be very large even for larger PBHs.

\subsection{$ T_r$ sets a cut-off}

Our setup is such that the universe after inflation is once dominated by an oscillating scalar field and then reheated to the temperature $ T_r$\,.
We neglect any PBHs produced before reheating (i.e., during the early matter domination) and even those produced after reheating if it happens on sub-horizon scales.
This gives us conservative PBH constraints. 
In Ref.~\cite{Khlopov:1980mg}, the authors discussed PBH formation in matter-dominant universe. Assuming a spherical collapse of a dense region into a PBH, they obtain the result that
more PBHs tend to be produced. This would be reasonable since the pressure $P=0$.
However, they did not consider any non spherical effects, which cause the
non-spherical morphologies to evolve during the  collapsing phase, which would have prevented a further collapse. 
To get a rigorous bound on the PBH spectrum for the early matter phase, we need numerical simulations to obtain the correct criterion 
for matter collapse in matter domination analogous to the radiation domination case that $1/3  < \delta <  1$ for matter to  collapse into a PBH.

\begin{figure}
\centering\includegraphics[scale=0.3]{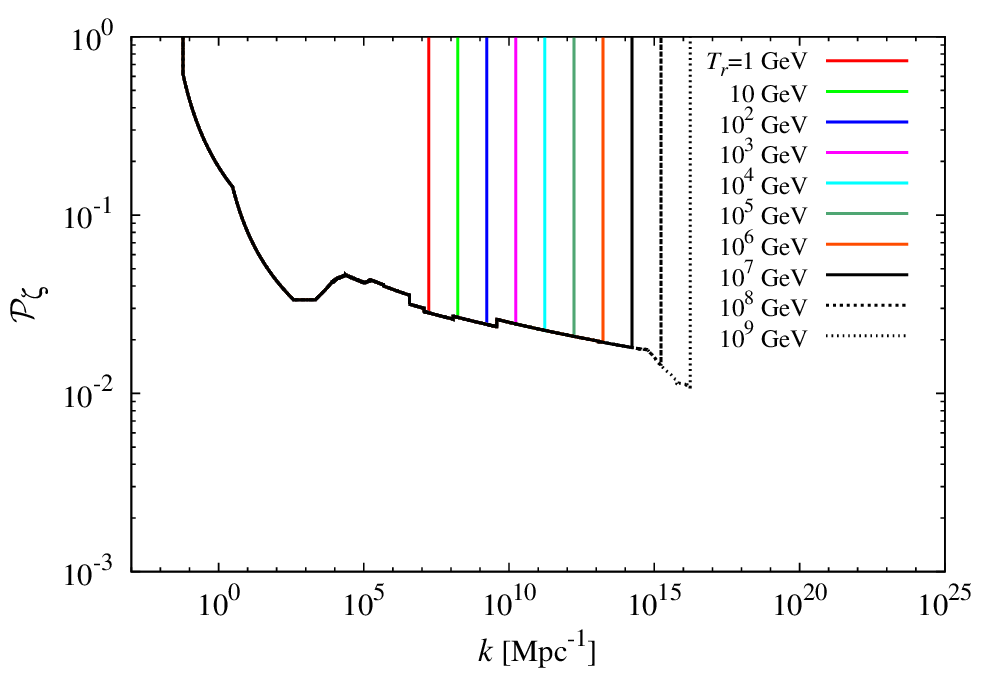}
\caption{The PBH bound on the primordial spectrum for various cut-off masses, as defined in the figure legend.}
\label{ps_rt}
\end{figure}

In fact, in this treatment there arises a cut-off PBH mass determined by the reheat temperature:
\begin{equation}
M_\mathrm{BH,co}
\approx
  0.946 \times 10^{32}\,\mathrm g\,
  \left(\frac{T_r}{1\,\mathrm{GeV}}\right)^{-2}\,,
\end{equation}
which corresponds to a cut-off wavenumber
\begin{equation}
k_\mathrm{co}
\approx
  1.71 \times 10^7\,\mathrm{Mpc}^{-1}\,
  \left(\frac{T_r}{1\,\mathrm{GeV}}\right)\,.
\end{equation}
The scalar perturbations with $ k > k_\mathrm{co}(T_r) $ 
may not constrained by PBHs, see \fig{ps_rt}.

\section{The spectra of induced gravitational waves}
\label{sec:MD}
We follow the analysis of Ref.~\cite{Ananda:2006af}
with much of the notation of Ref.~\cite{Baumann:2007zm}. 
The spectrum of induced gravitational waves generated during an early matter phase was 
first calculated in Ref.~\cite{Assadullahi:2009jc} and is given as
\be
\mathcal{P}_h(k)=\frac{2}{x^3}\(\frac{40}{3}\)^2\int_0^\infty dv\int_{|v-1|}^{|v+1|}dy\frac{v^2}{y^2}(1-\mu^2)^2\mathcal{P}_\zeta(ky)\mathcal{P}_\zeta(kv)I_{MDS}^2
\ee
where $x=k\tau$, $k$ is the scale, $\tau$ is conformal time, $v=\tilde{k}/k$ the ratio of the incoming to outgoing scales, 
$y=\sqrt{1+v^2-2v\mu}$, $\mu$ is the cosine of the angle between the incoming and outgoing scales, $\mathcal{P}_\zeta$ is the spectrum of 
primordial fluctuations, in this case generated during inflation, and $I_{MDS}$ is the time integral given as
\be
I_{MDS}=\frac{3x\cos(x)-3\sin(x)+x^3}{ x^{3/2}}~.
\ee
We have taken the lower limit to be $x=0$, and terminated the integration at $x=k\tau_r$, 
where $\tau_r$ is the conformal time at the end of the reheating era.

The spectrum for induced gravitational waves during radiation domination is given by \cite{Ananda:2006af,Baumann:2007zm}
\be
\mathcal{P}_h(k)=\frac{k^2}{x^2}\int_0^\infty{}dv\int_{|v-1|}^{|v+1|}dy\frac{v^2}{y^2}(1-\mu^2)^2\mathcal{P}_\zeta(kv)\mathcal{P}_\zeta(ky)\tilde{I}_1\tilde{I}_2 .
\label{eq:Ph}
\ee
where the time integrals, $\tilde{I}_1$ and $\tilde{I}_2$, used in this paper
are derived in \cite{Alabidi:2012ex} and given in appendix \ref{app:rad-time}.

\subsection{Analytical Estimate for a Flat Spectrum}

In this section we assume a flat spectrum and set $\mathcal{P}_\zeta(k)=\Delta_R\simeq10^{-9}$. Then the spectrum can be written as:

\be
\mathcal{P}_h(k)=2\(\frac{40}{3}\)^2\Delta_R^2\int dv\int dy(1-\mu^2)^2\(\frac{v}{y}\)^2\frac{I_{MD}^2}{x^3}
\ee
consider the term $I_{MDS}/x^{3/2}$:
\be
\frac{I_{MD}}{x^{3/2}}=\left[1+\frac{3}{x^3}\(x\cos(x)-\sin(x)\)\right]\equiv I_{MDS}
\ee
and it is clear that for $x\gg1$ $I_{MDS}$ approaches a constant. Substituting this into our equation for the spectrum we have
\be
\mathcal{P}_h(k)=2\(\frac{40}{3}\)^2\Delta_R^2\int dv\int dy(1-\mu^2)^2\(\frac{v}{y}\)^2I_{MDS}^2
\ee
recall that $x=k\tau$, where $\tau$ is the limit of our time integral which we take to be the end of the reheating phase $\tau_r=2/k_r$. Therefore $x_r$ is much greater than $1$ for most of the scales we consider. Hence we can pull $I_{MDS}$ out of the integral
\be
\mathcal{P}_h(k)=2\(\frac{40}{3}\)^2\Delta_R^2I_{MDS}^2\int dv\int dy(1-\mu^2)^2\(\frac{v}{y}\)^2
\ee
and the integrals can be performed analytically. We find that the analytical equation is compatible with with numerical calculation for a flat spectrum.

For scales $k_r<k<k_{max}$ the spectrum is then
\bea
\label{eq:P_an}
\frac{\mathcal{P}_h}{\Delta_R(k)^2}&\approx&2\(\frac{40}{3}\)^2\left(\frac{16\, k}{35\, \mathrm{k_{max}}} + \frac{16\, \mathrm{k_{max}}}{15\, k} - \frac{4\, {\mathrm{k_r}}^4}{15\, k^4} + \frac{8\, {\mathrm{k_r}}^6}{105\, k^6}\right)
\eea
where we have taken the upper limit on $v$, $v_{max}=k_{max}/k$ and the lower limit to be $v_{min}=k_{min}/k$, and we took $k_{min}=k_r$ 
where $k_r$ is the scale that re-entered the horizon at the end of the matter era, we have also taken $I_{MDS}\approx1$. By only considering scales which re-enter the horizon near $k\sim k_r$, those whose amplitudes have grown the most, \eq{eq:P_an} can be reduced
to $\sim 356 (16k_{max}/(15k))$ which for $k_{max}=k_{NL}\sim 141 k_r$ and $k\sim k_r$ is $10^5$. Taking instead $k_{max}=10^3k_r$ leads to a spectrum
maximum of $\approx 10^6$ as is confirmed in the full numerical calculation shown in \fig{fig:flat}.
\subsection{The evolution  of the tensor mode}

Defining $v_\bk=ah_\bk$, the equation of motion for the tensor modes is given as
\be
v^{''}_\bk+\left(k^2-\frac{a''}{a}\right)=aS_\bk
\label{eq:v}
\ee
and can be solved approximately for the full evolution of the universe. Using step and boxcar functions, the source term $aS_\bk$ can be  written out as
\be
a\mathcal{S}_\bk\propto k^2\tau^2\(\theta(\tau_r-\tau)+\frac{\tau_r^4}{\tau^4}\Pi_{\tau_r\tau_{eq}}+\left(\frac{\tau_r}{\tau_{eq}}\right)^4\theta(\tau-\tau_{eq})\)
\label{eq:sudden}
\ee
where we have taken $\mathcal{S}\propto\tau^{-3}$ during radiation domination, $\theta$ is the heaviside step function and $\Pi_{\tau_r\tau_{eq}}=\theta(\tau-\tau_r)-\theta(\tau-\tau_{eq})$ is the boxcar function. The scale factor can be written out in a similar fashion
\be
a\propto\tau^2\theta(\tau_r-\tau)+\tau\tau_r\Pi_{\tau_r\tau_e}+\(\frac{\tau_r}{\tau_{eq}}\)\tau^2\theta(\tau-\tau_{eq})
\ee

\begin{figure}
\centering
\includegraphics[scale=0.3]{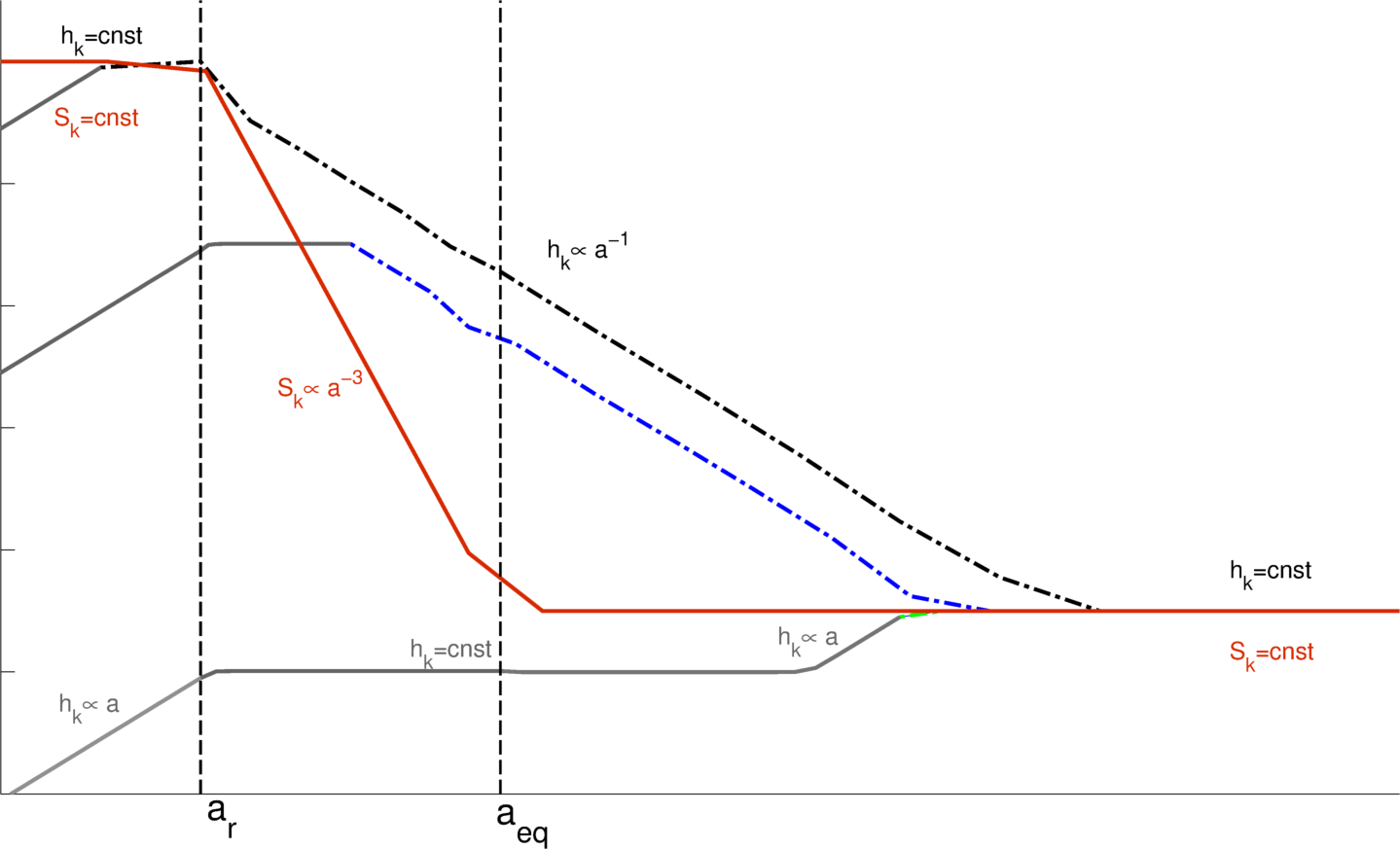}
\caption{A schematic of the evolution of tensor modes on different scales with respect to the logarithm of the scale factor. The solid red line depicts the evolution of the source term between epochs.
$a_r$ is the scale factor at reheating and $a_{eq}$ is the scale factor at radiation-matter equality. Grey lines represent super-horizon modes. The black dash-dot line represents the evolution of a mode which enters during EMD. The blue dash-dot line represents the amplitude of the mode which enters during RD and the green line, which is barely visible at the right hand side of the plot, represents the mode which enters during the current epoch (assuming no acceleration).} \label{fig:sub_sup}
\end{figure}

For sub horizon modes, $k\gg1/\tau$, we obtain the solution plotted black in \fig{fig:sub_sup}. 
In this scenario, inflation gives way to an early phase of matter domination which ends 
when $\tau=\tau_r$ and is followed by a phase of radiation domination that is overtaken by matter at $\tau=\tau_{eq}$. 
The source term is at first constant, then when $\tau_r<\tau<\tau_{eq}$ it decays at a rate $\propto\tau^{-3}$ and 
constant again for $\tau>\tau_{eq}$. The amplitude of the sub-horizon tensor modes which re-enter the horizon during early 
matter domination is held at a constant until $\tau_r$, 
when begins to freely propagate and decay at a rate $\propto a^{-1}$, 
until it becomes equal to the source term and held at a constant value. The superhorizon modes grow until $\tau=\tau_r$, are held at a constant between $\tau_r<\tau<\tau_{eq}$ and grow again 
for $\tau>\tau_{eq}$. 


\subsubsection{The accuracy of the sudden transition approximation}
\begin{figure}
\centering\includegraphics[scale=0.3]{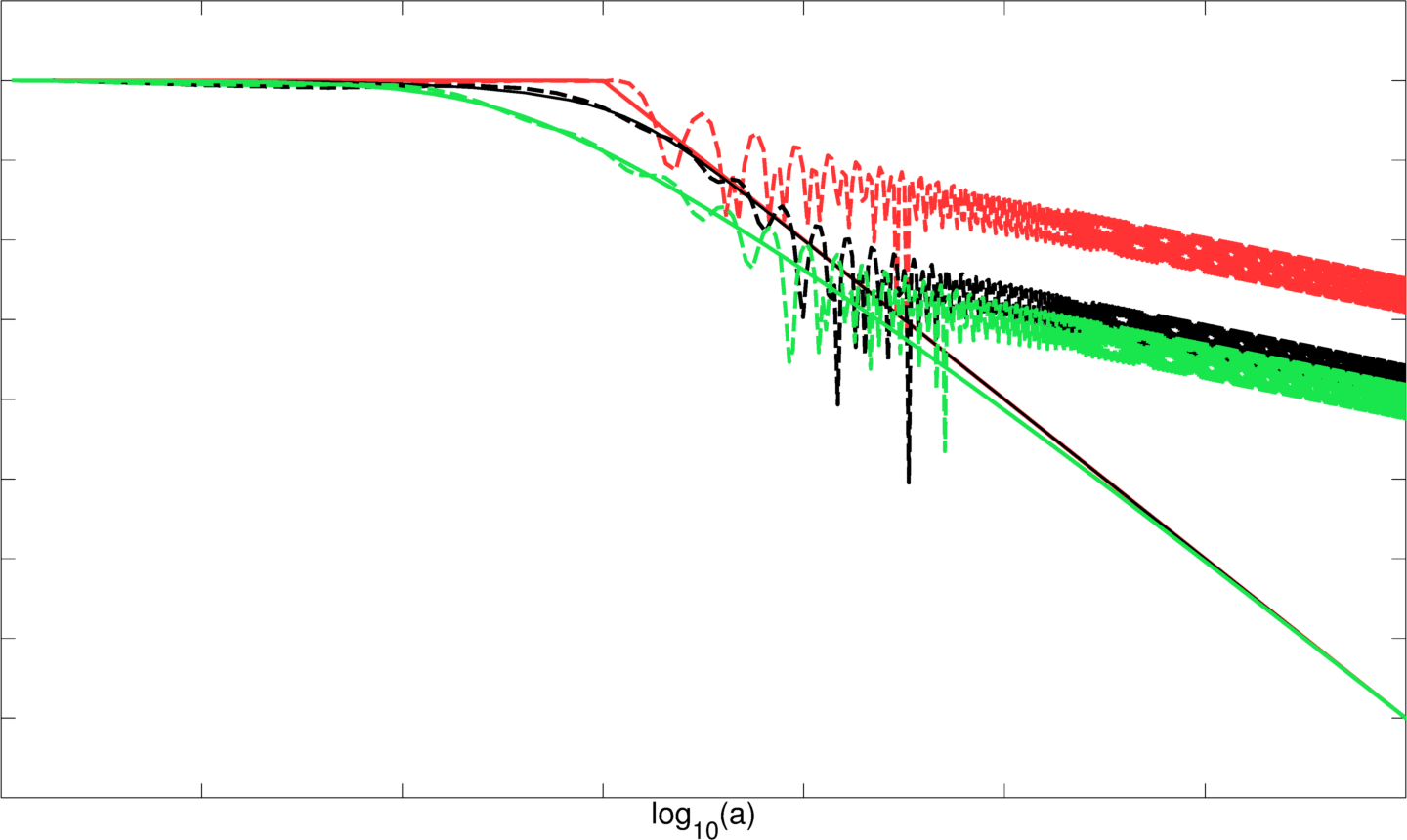}
\caption{The solid lines represent the source term, the red is the sudden
transition approximation \eq{eq:sudden}, the green is the smooth turnover with  $n=1$, and the black is $n=4$. The dashed lines represent the tensor modes. As we can see the tensor modes do approach
the freely oscillating limit, but there is some loss of amplitude with respect to the sudden transition approximation}
\label{fig:ode_smooth}
\end{figure}

Throughout this paper we have utilised the sudden transition approximation between an early matter phase and radiation. In this section we investigate the effect a smoother turnover has on the tensor modes generated during the early matter phase. For this we make the following approximations for the scale factor and source term
\bea
a&=&2 a_r\(\frac{\tau}{\tau_r}\)^2\frac{1}{(1+(\tau/\tau_r)^n)^{1/n}}\nonumber\\
S&=&\frac{k^2}{1+(\tau/\tau_r)^4}
\eea
where $n$ is an integer. We plot the results for $n=1$, $n=8$ and the sudden transition approximation in \fig{fig:ode_smooth}. As is clear from the figure, in all cases the tensor modes approach the freely oscillating stage during radiation domination, however, the smooth turnover results in a smaller amplitude of an order of magnitude. We also note that $n=8$ is very close to the sudden transition approximation, and that $n=1$ is less than an order of magnitude smaller that it. This phenomenon requires further investigation.

\subsection{Transfer Function}

Detectors of gravitational waves will place a bound on the energy density of gravitational waves, defined
as ~\cite{Maggiore:2000gv}:
\be
\Omega_{\rm GW}(f)=\frac{1}{\rho_c}\frac{d\rho_{\rm GW}}{d\ln{}f}
\ee
where $f$ is the frequency, $\rho_c$ is the critical energy density defining
the coasting solution of the Friedman equation and $\rho_{\rm GW}$ is the energy
density of gravitational waves. This is related to the primordial spectrum via a transfer function, $\Omega=t^2(k,\tau)\mathcal{P}_h(k)$.
Scales that re-enter the horizon during a period of early matter domination experience a constant source term, 
and the amplitude of the tensor mode is kept at it's super horizon value. However once the universe enters 
the radiation epoch the source term decays as does the amplitude of the tensor modes, as is depicted in 
\fig{fig:sub_sup}.

In Ref.~\cite{Baumann:2007zm} they show that for  scales smaller than some critical scale, 
which our scales of interest are, the transfer function is
\be
t(k,\tau)=\frac{a_k}{a(\tau)}
\ee
where $a_k$ is the scale factor when the scale $k$ enters the horizon. 
To be precise, it is the scale factor at the time when the source term at that scale begins to decay. 
That is, for scales that enter the horizon during early matter domination, $a_k=a_r$. Our transfer function is then
\bea
t(k,\tau)&=&\frac{a_r}{a(\tau)}\nonumber\\
&=&\frac{a_r}{a_{eq}}\frac{a_{eq}}{a(\tau)}\nonumber\\
&=&\frac{a_{eq}}{a_r}\frac{k_{eq}}{k_r}
\eea
where subscript $eq$ is that of radiation-matter equality and the relative energy of scalar-induced gravitational waves is
\bea
\Omega_{GW}(k,\tau)&=&\frac{a(\tau)k^2}{a_{eq}k_{eq}^2}t^2(k,\tau)\mathcal{P}_h\nonumber\\
&=&\frac{1}{(1+z_{eq})}\(\frac{k}{k_r}\)^2\mathcal{P}_h
\eea
where $z$ is the redshift. For scales which re-enter the horizon during radiation domination, the relative energy of induced gravitational waves is
\be
\Omega_{GW}=\frac{1}{1+z_{eq}}\mathcal{P}_{h}(k)~.
\label{eq:omega-rad}
\ee

\subsection{Full numeric results for a flat spectrum}

To get the full spectrum of induced gravitational waves for an early matter phase followed by a radiation phase we 
evaluate
\bea
\mathcal{P}_h(k)&=&\(\int_0^{x_r} F(v,y,\tau_1)d\tau_1+\int_{x_r}^x F(v,y,\tau_2)d\tau_2\)\nonumber\\
&=&\int_0^{x_r} F(v,y,\tau_1)F(v,y,\tau_2)d\tau_1 d\tau_2+\nonumber\\
&&\(\int_0^{x_r}F(v,y,\tau_1)d\tau_1\int_{x_r}^xF(v,y,\tau_2)d\tau_2+\int_0^{x_r}F(v,y,\tau_2)d\tau_2\int_{x_r}^xF(v,y,\tau_1)d\tau_1\)\nonumber\\
&&+\int_{x_r}^x F(v,y,\tau_2)F(v,y,\tau_1)d\tau_1 d\tau_2\nonumber\\
&=&\mathcal{P}_{h,matter}+C_{h,cross}+\mathcal{P}_{h,radiation}\nonumber\\
&\approx &\mathcal{P}_{h,matter}+\mathcal{P}_{h,radiation}
\eea
where $F(v,y,\tau)$ is the integral over $v,y$ in \eq{eq:Ph}, and we drop the cross terms arising from $<\Phi_{matter}\Phi_{radiation}>$ in the last line. This approximation is reasonable since the cross terms are only of significance at $k\sim k_r$.

\begin{figure}
\centering\includegraphics[width=5in,totalheight=3in]{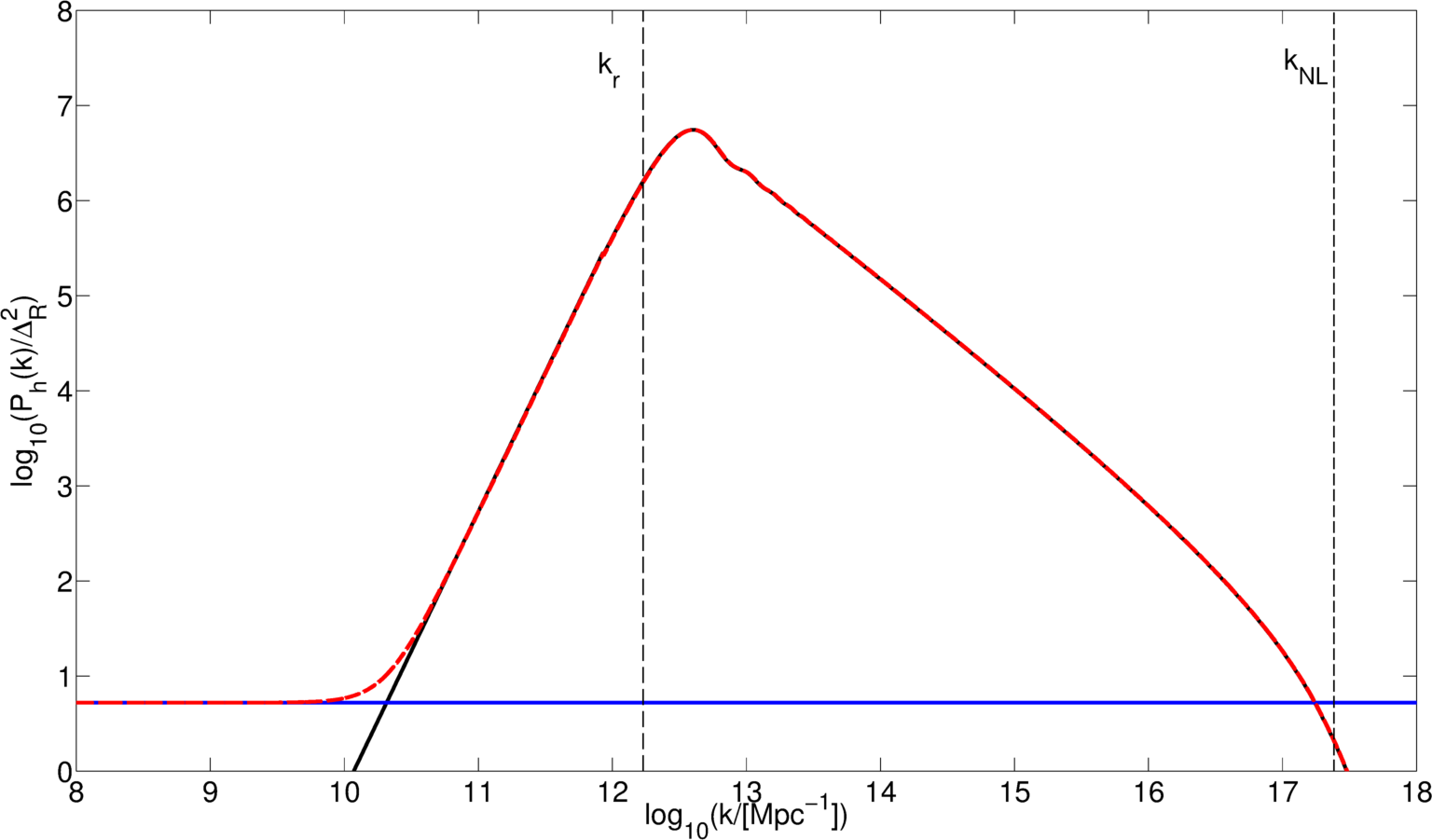}
\caption{We plot the spectrum of induced gravitational waves for a flat primordial spectrum with $T_r=10^9\rm{GeV}$. 
The spectrum for scales that re-enter the horizon deep in the radiation era have a flat spectrum, due to the fact that the source term 
is a decaying function and thus the modes oscillate freely, and is represented by the solid blue line. 
The solid black line represents the modes that re-enter the horizon during the early matter phase ($k>k_R$). The 
The red dashed line is the complete spectrum, assuming an early matter phase followed by a phase of radiation domination. 
The spectrum for $k>k_{NL}$ behaves as $\mathcal{P}_h\propto 1/k^4$, as $\mathcal{P}_h\propto 1/k$ for $k_r<k<k_{NL}$, as $\mathcal{P}_h\propto k^3$ for $k\lesssim k_r$
and as $\mathcal{P}_h\sim \rm{constant}$ for $k\ll k_r$.
One can think of this as follows: modes that re-enter the horizon during the radiation phase but with $k\sim k_r$,
 i.e. near the EMD phase, will interact with modes that re-entered during EMD and hence their behaviour/characteristics 
are modified from the instant reheating scenario. We have utilised a simplified analysis, in that $\mathcal{P}_h(k)=\mathcal{P}_{h_{matter}}(k)+\mathcal{P}_{h_{rad}}(k)$.}
\label{fig:flat}
\end{figure}

Figure \ref{fig:flat} is a depiction of the spectrum of induced gravitational waves arising from 
a flat primordial spectrum of density perturbations; $n_s=1$. 
This figure was generated mainly for illustrative purposes, and as such we have chosen $k_{NL}$ 
to be $10^3$ times as large as the value calculated using \eq{eq:knl}. 
We assume that modes with $k>k_{NL}$ do not experience the constant source term. 
Scales which are still super-horizon at the end of the early-matter phase have a spectrum $\mathcal{P}_h\propto k^3$ 
and therefore become rapidly smaller than the spectrum generated by the pure radiation source term. 
That means that for modes which enter the horizon soon after $\tau_r$ we need only consider the convolution 
of modes with those that enter during the radiation era.

\begin{figure}
\centering\includegraphics[width=5in,totalheight=3in]{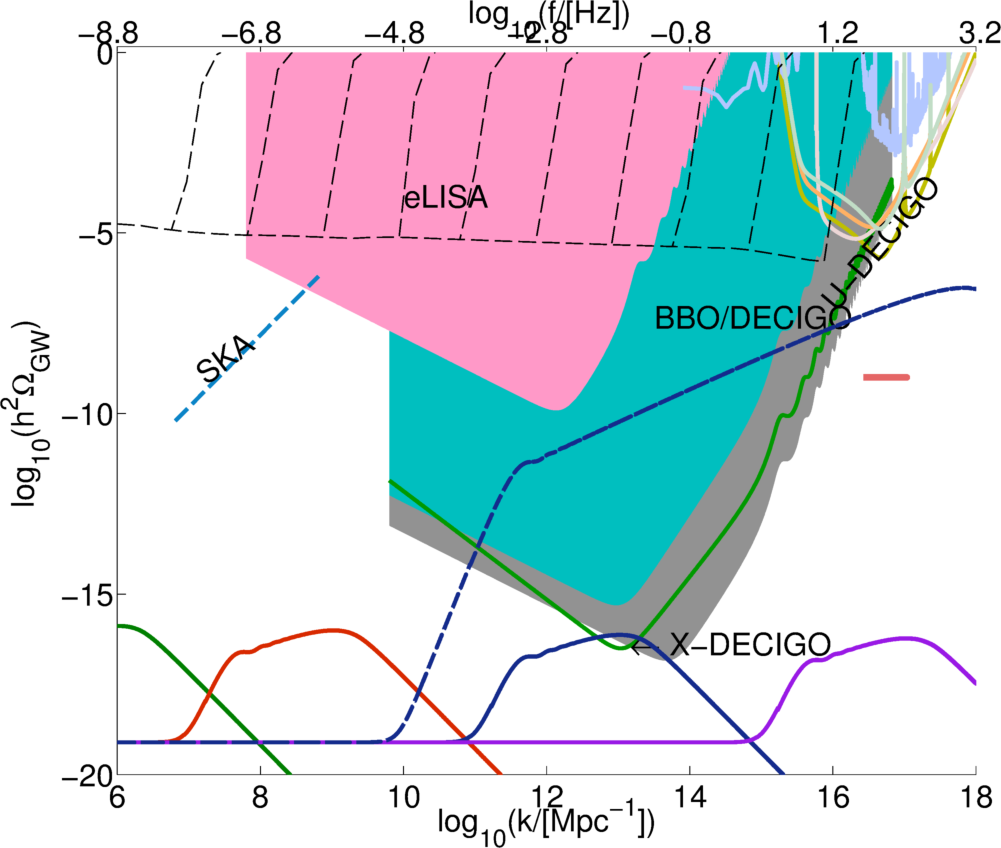}
\caption{We plot the spectrum of induced gravitational waves for a 
flat primordial spectrum with various reheat temperatures and the cutoff scale $k_{max}=k_{NL}$. 
The black dashed lines correspond to the PBH bound assuming (from left pseudo-vertical line to right pseudo-vertical line) $T_r=1\rm{GeV},10\rm{GeV},10^2\rm{GeV},10^3\rm{GeV},10^4\rm{GeV},10^5\rm{GeV},10^6\rm{GeV},10^7\rm{GeV}, 10^8\rm{GeV}$ and $10^9\rm{GeV}$.
The cluster of solid lines in the top right corner are the sensitivity ranges of ground based detectors, 
LIG0 S5 and S6 \cite{ligo}, 
and KAGRA \cite{kagra}, while the thick horizontal salmon pink line is the forecast sensitivity of Advanced LIGO \cite{adv_ligo,Abbott:2009ws}.
Also shown is the sensitivity limit of the Square Kilometre Array (SKA) \cite{Jenet, Hobbs, Yardley, PPTA}.
The green, red, blue and purple solid lines correspond to taking $T_r=1\rm{MeV}$, $1\rm{GeV}$,
$10^4\rm{GeV}$, and $10^8\rm{GeV}$. The blue dashed line is the spectrum for $T_r=10^4\rm{GeV}$ without terminating
at $k_{NL}$. It is interesting to note that even a flat primordial spectrum can 
lead to a spectrum of induced gravitational waves detectable by cross-correlated DECIGO}
\label{fig:flat_sens}
\end{figure}

We present the results for a flat spectrum at various reheat temperatures in \fig{fig:flat_sens},
 where we have taken the limits on $v$ to be $v_{min}=k_r/k$ and  $v_{max}=k_{NL}/k$ with the latter upper bound accounting for the non-linear cutoff. We could have modified the calculation and checked for each $v$ and $y$ that $\tilde{k}<k_{NL}$ and $|\tilde{\mathbf{k}}-\mathbf{k}|<k_{NL}$, however simply modifying the limits of $v$ has the same effect.


\section{The Models of Inflation}\label{sec:models}
\begin{figure}
\centering\includegraphics[width=\linewidth,totalheight=2in]{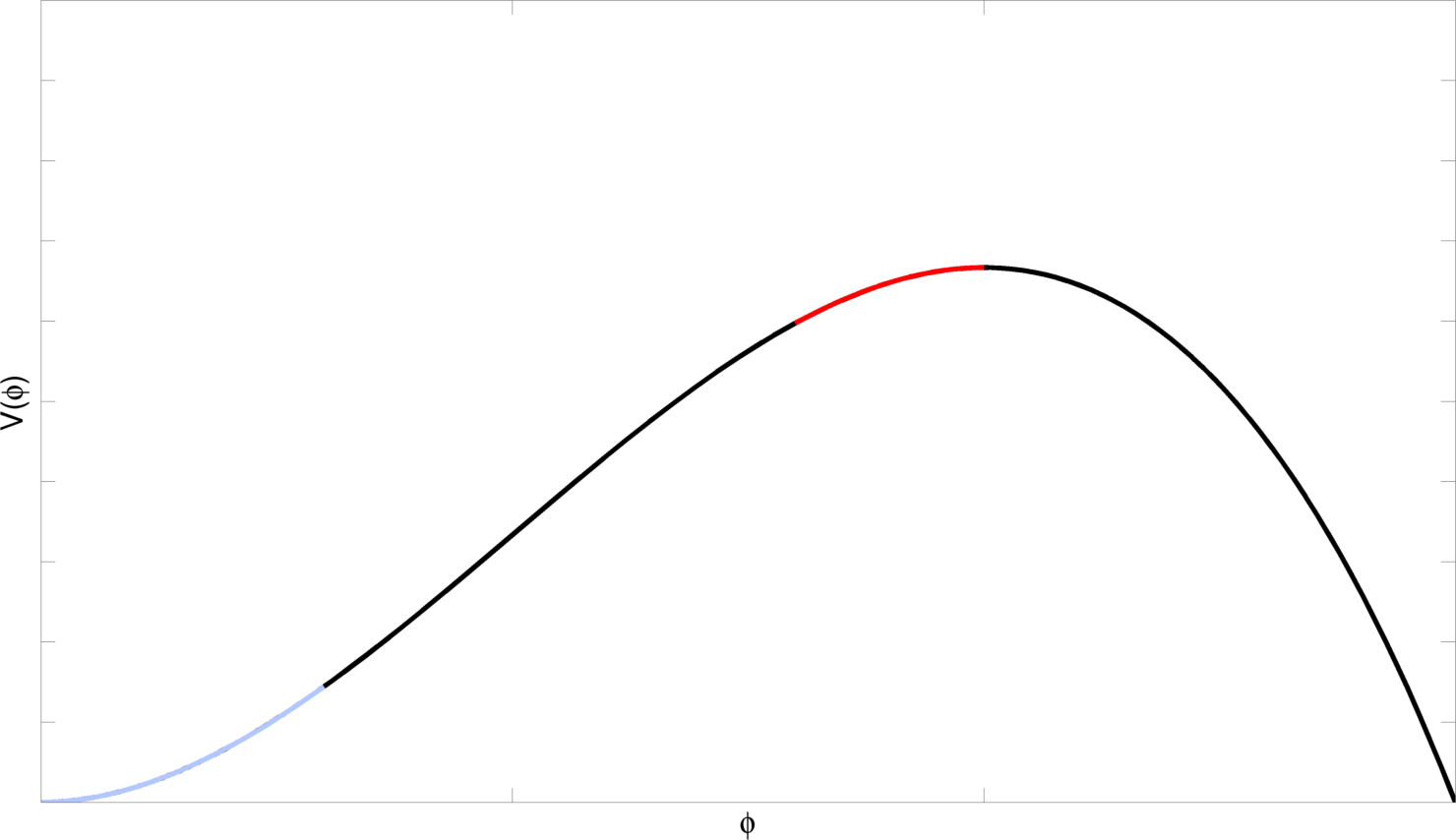}
\caption{An illustration of the Hilltop-type and running models. In our scenario scales of cosmological
interest during the hilltop regime (indicated with red), and the end of inflation occurs once the inflaton
has reached a flatter region of the potential (indicated with blue).}
\label{fig:model}
\end{figure}

The spectrum of induced gravitational waves is directly proportional to the square of the primordial spectrum 
and is therefore clear that for an enhanced spectrum of induced gravitational waves one needs an enhanced primordial 
spectrum. Since at the pivot scale the spectrum is tightly constrained by CMB data we need to go beyond this and 
consider models which enhance the spectrum on small scales.
Phenomenologically, the two models of inflation that exhibit this property \cite{Kohri:2007gq, Drees:2011yz} are the running 
mass model and the hilltop model, depicted in \fig{fig:model}.

\subsection{The Hilltop type model}
Identified in Ref.~\cite{Kohri:2007gq} as the phenomenological form necessary for PBH formation this model has the potential \cite{Kohri:2007qn,Alabidi:2009bk}:
\be\label{hilltop-potential}
V=V_0\(1+\eta_p\varphi^p-\eta_q\varphi^q\)
\ee
where the coupling terms $\eta_p$ and $\eta_q$ are less than $1$ and $p<q$ to get the shape in \fig{fig:model}. 
Certain realisations in super gravity can be found; see for example Refs.~\cite{Allahverdi:2006cx,Kohri:2007gq,Lin:2008ys,Lin:2009yt,BenDayan:2009kv,Kohri:2010sj,Hotchkiss:2011gz}.

This model is very compatible with WMAP data, and as such many of the model's 
terms are not ruled out, but in this analysis we add the extra requirement that the model 
is maximised at small scales and yet still remain within the PBH bound. We also take the basic number of $e-$folds to be $N=56$.
Parameter selection criteria is explained in more detail in Ref.~\cite{Alabidi:2012ex}.

\subsection{The Running Mass Model}
This model is the basic $\phi^2$ model with a varying mass term that arises in taking renormalised group equations, and is given as
\cite{Stewart:1996ey,Covi:1998jp, Covi:1998mb, Lyth:2000qp,Leach:2000ea,Covi:2000qx, Covi:2002th,Covi:2004tp,Bugaev:2009kq}
\be
\frac{V}{V_0}=1-\frac{B_0}{2}\vp^2+\frac{A\vp^2}{2(1+\alpha\ln(\vp))^2}~.
\ee
In this case we select parameters which satisfy $n_s=0.96$, $n_s'=0.0039$ and $n_s'=0.0043$, which for $T_r>10^9\rm{GeV}$ are terminated
at $N=$ and $N=$ respectively. 
\section{Results}
\label{sec:results}

\begin{figure}
\centering\includegraphics[scale=0.4]{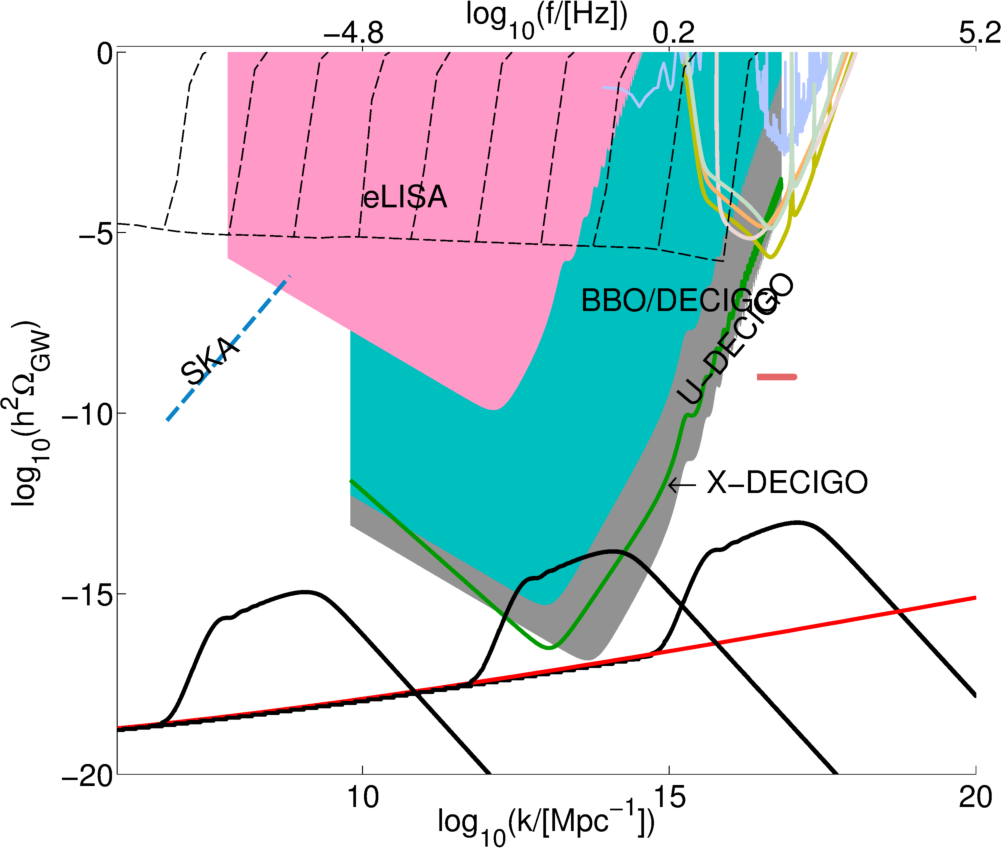}
\caption{The spectra of induced gravitational waves for the Hilltop model with $p=2$ and $q=3$, with $k_{max}=k_{NL}$. 
The black dashed lines and the cluster of solid lines in the right hand corner are defined in \fig{fig:flat_sens}. 
The red line represents the induced gravitational wave spectrum for $T_r\gg10^{10} \rm{GeV}$ and $N=55$ $e-$folds. 
From right to left the black solid lines correspond to the induced gravitational wave spectrum for reheat temperatures of 
$T_r=10^8\rm{GeV}$, $T_r=10^5\rm{GeV}$, and $T_r=1\rm{GeV}$.
}
\label{fig:23}
\end{figure}

\begin{figure}
\centering\includegraphics[scale=0.4]{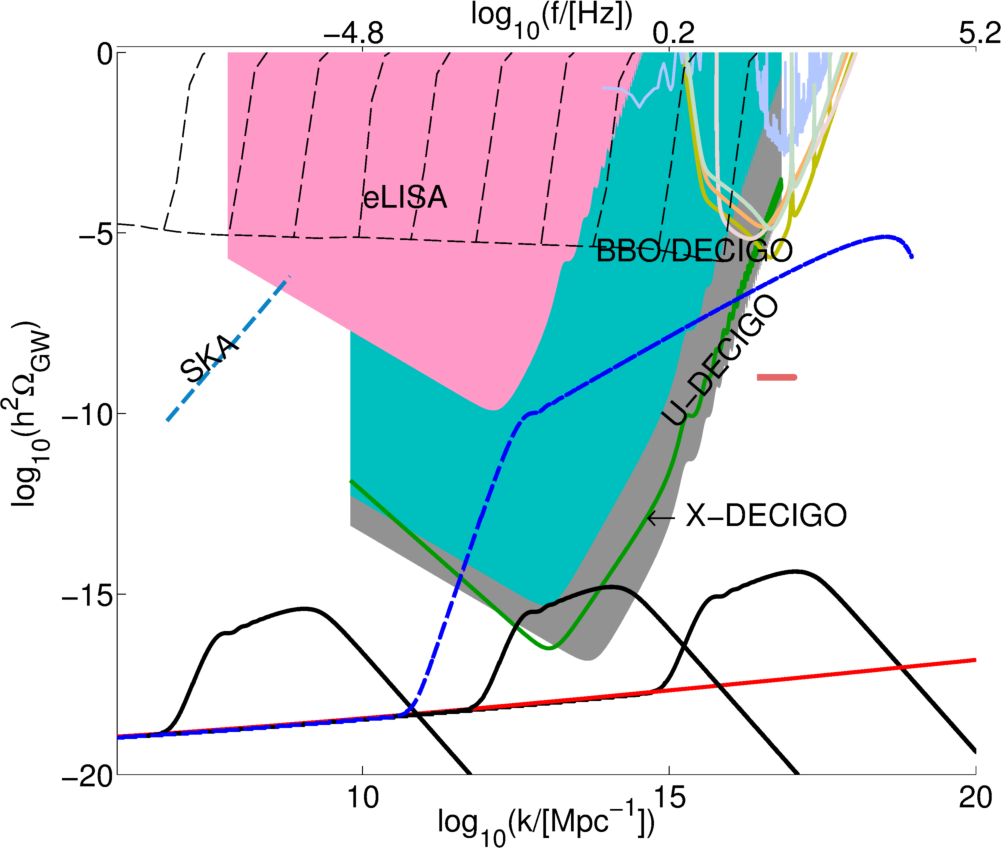}
\caption{The spectra of induced gravitational waves for the Hilltop model $p=2$ and $q=4$, with $k_{max}=k_{NL}$. 
The black dashed lines and the cluster of solid lines in the right hand corner are defined in \fig{fig:flat_sens}. 
The red line represents the induced gravitational wave spectrum for $T_r\gg10^{10} \rm{GeV}$ and $N=55$ $e-$folds. 
From right to left, the black solid lines correspond to the induced gravitational wave spectrum for $T_r=10^8\rm{GeV}$, $T_r=10^5\rm{GeV}$, 
and $T_r=1\rm{GeV}$. For illustration, we include the spectrum for a reheat temperature of $T_r=10^5\rm{GeV}$
and integrated up to the maximum scale instead of the non-linear cutoff (dashed blue line).}
\label{fig:24}
\end{figure}

\begin{figure}
\centering\includegraphics[scale=0.4]{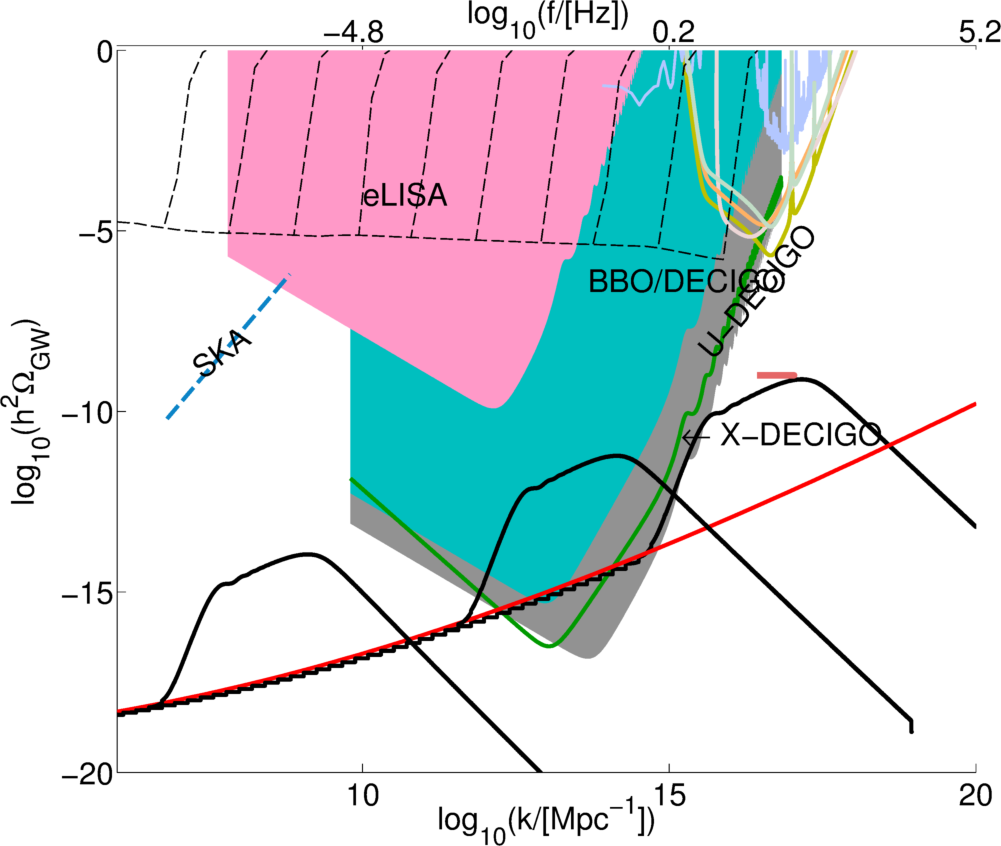}
\caption{The spectra of induced gravitational waves for the Hilltop model $p=2$ and $q=2.3$, with $k_{max}=k_{NL}$.  
The red line represents the induced gravitational wave spectrum $T_r\gg10^{10} \rm{GeV}$ and $N=55$ $e-$folds. From right to left, the black solid lines correspond 
to the induced gravitational wave spectrum for $T_r=10^8\rm{GeV}$, $T_r=10^5\rm{GeV}$, and $T_r=1\rm{GeV}$. 
The black dashed lines and the cluster of solid lines in the right hand corner are defined in \fig{fig:flat_sens}. }
\label{fig:223}
\end{figure}

\begin{figure}
\centering\includegraphics[scale=0.4]{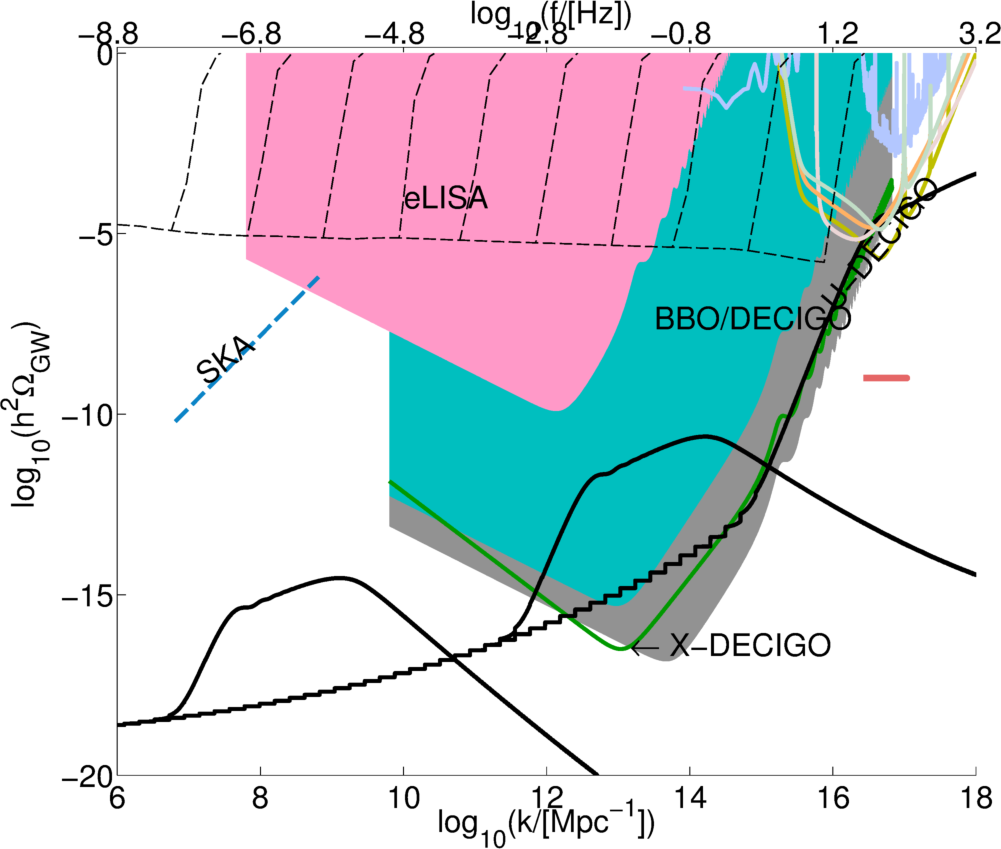}
\caption{The spectra of induced gravitational waves for the running mass model satisfying $n_s'=0.0039$ and $N=57$ , with $k_{max}=k_{NL}$. 
From right to left, the black solid lines correspond to the induced gravitational wave spectrum $T_r=1\rm{GeV}$,$T_r=1\times10^5\rm{GeV}$, and $T_r=10^9\rm{GeV}$. 
The black dashed lines and the cluster of solid lines in the right hand corner are defined in \fig{fig:flat_sens}. }
\label{fig:rmm1}
\end{figure}

\begin{figure}
\centering\includegraphics[scale=0.4]{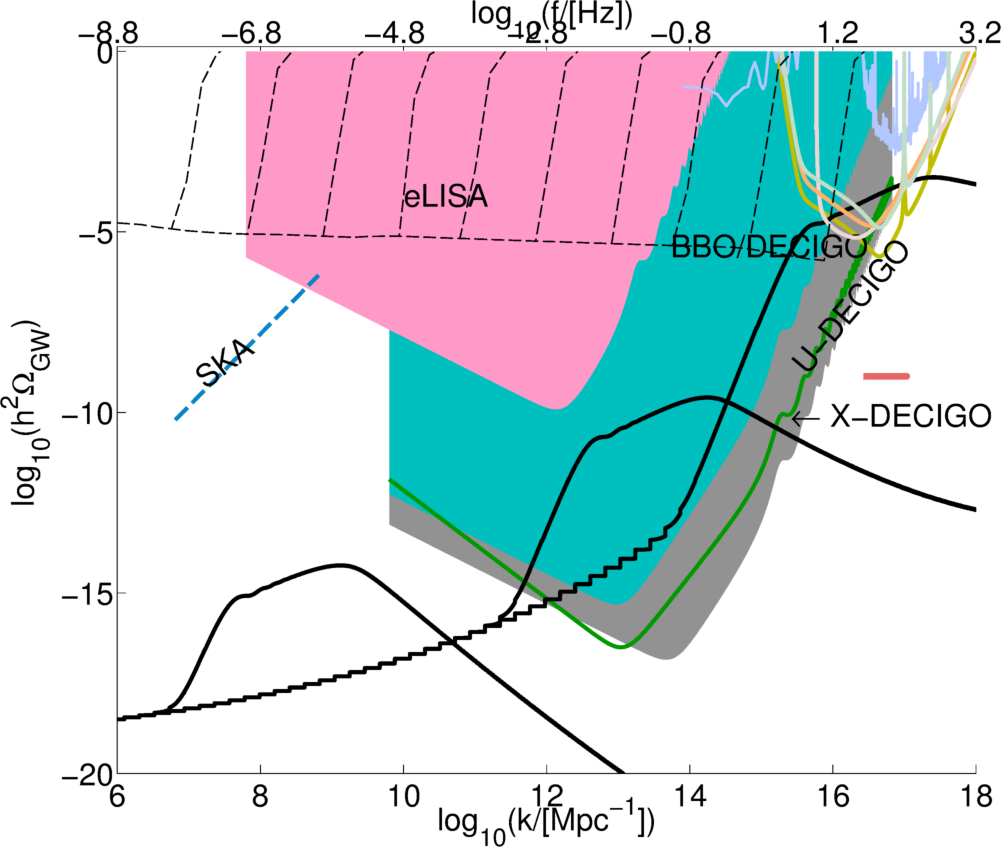}
\caption{The spectra of induced gravitational waves for the running mass model satisfying $n_s'=0.0043$ and $N=57$, with $k_{max}=k_{NL}$. 
From right to left, the black solid lines correspond to the induced gravitational wave spectrum $T_r=1\rm{GeV}$,$T_r=1\times10^5\rm{GeV}$, and $T_r=10^8\rm{GeV}$.T
The black dashed lines and the cluster of solid lines in the right hand corner are defined in \fig{fig:flat_sens}. }
\label{fig:rmm2}
\end{figure}

\begin{figure}
\centering\includegraphics[scale=0.4]{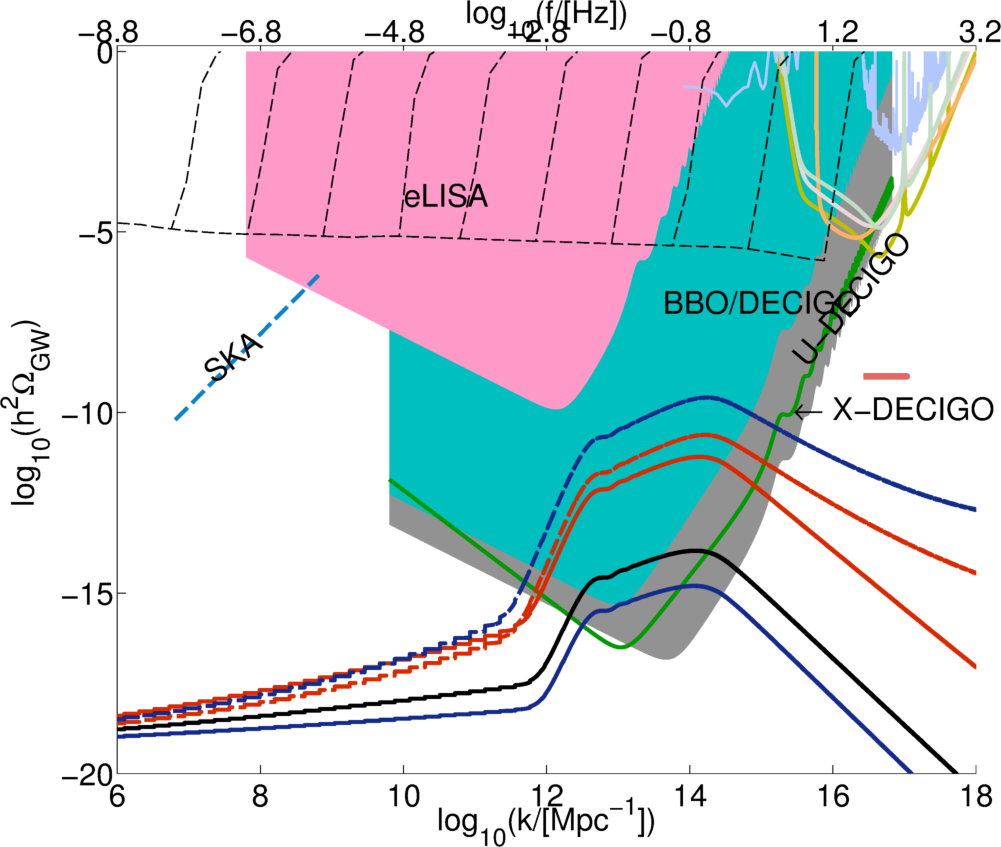}
\caption{The spectra of induced gravitational waves for all the models plotted in the previous 
figures assuming a reheat temperature of $T_r=10^5\rm{GeV}$, with $k_{max}=k_{NL}$.
The solid lines are hilltop with $p=2$ $q=2.3$ (red), $p=2$ $q=3$ (black) and $p=2$ $q=4$ (blue); 
the dashed lines are the running mass models with $n_s'=0.0043$ (blue) and $n_s'=0.0039$ (red).
The black dashed lines and the cluster of solid lines in the right hand corner are defined in \fig{fig:flat_sens}. }
\label{fig:15}
\end{figure}

We plot results for the hilltop model for $p=2$ and $q=2.3,3,4$, and for the running mass models which satisfy $n_s=0.96$ and $n_s'<0.0062$ 
with $N\sim57$. These are plotted in
Figures.~\ref{fig:23},\ref{fig:24}, \ref{fig:223}, \ref{fig:rmm1}, \ref{fig:rmm2} and \ref{fig:15} 
for a range of reheat temperatures $1\rm{GeV}<T_r<10^9\rm{GeV}$ \footnote{There is an error in our previous paper, a missing factor of $4$ in the $f$ function which
 means that our previous results are $4^2$ times smaller than they should be, this is corrected for here.}. 
Each reheat temperature modifies the maximum allowed number of $e-$folds $N_{max}$ 
and therefore we have integrated only up to $k_{max}=k_{pivot}e^{N_{max}}$, 
except for in the case when $k_{max}>k_{NL}$ where we only integrate up to $k_{NL}$. In the final figure, 
\fig{fig:15} we have plotted the results of the hilltop and running mass models for a reheat temperature of $T_r=10^6\rm{GeV}$.

In our previous paper Ref.~\cite{Alabidi:2012ex}, we calculated the spectrum of 
Induced Gravitational Waves for the running mass models with large running $0.0067<n_s'<0.012$ 
which is no longer supported by the latest WMAP release \cite{Hinshaw:2012fq,Bennett:2012fp}. 
On a related note, to motivate $N<37$ by modifying the reheat temperature would require $T_r<1\rm{MeV}$ which is unsupported by theory.

We also find that for the running mass model satisfying $N=38$ $e-$folds, the corresponding induced gravitational waves spectra were not within the
sensitivity ranges of any of the experiments. However, if the $k_{NL}$ cutoff can be relaxed, the spectra may very well be within the ranges
of SKA and PULSAR.

\section{Discussion}
\label{sec:disc}

In this work we assumed a sudden transition between early matter domination and radiation domination. In the early universe however, the Hubble time can be of the order of the decay rate of matter to radiation. 
In this case modes re-entering the horizon towards the end of matter domination will also experience a decaying source term and the effect shown here may be  reduced, depending on how long the transition phase lasts, as is shown in \fig{fig:ode_smooth}. Therefore the results presented here are upper bounds on what the induced gravitational wave spectrum could be, with the actual spectrum possibly being only an order of magnitude or two smaller than what we calculated.

Therefore our conclusions depend on the condition that the tensor modes generated 
during an early matter phase survive the transition into radiation. Under this proviso, we have shown that assuming an early matter phase with $T_r=10^5\rm{GeV}$ results in a spectrum of 
induced gravitational waves with 
energy densities within the range of BBO/DECIGO and cross-correlated DECIGO \cite{Kudoh:2005as}. 
Since the assumption of an early matter phase truncates the PBH bound at a smaller value of $k$, 
we have shown that the running mass model generates induced gravitational waves detectable by 
ground based gravitational wave detectors such as LIGO, 
and KAGRA. 
This means that the RMM model with a running of $n_s'=0.0043$ and a reheat temperature 
of $10^8\rm{GeV}$ as well as the model with $n_s'=0.0039$ and $T_r=10^9\rm{GeV}$ can 
be ruled out since LIGO has failed to detect a gravitational signature.

\acknowledgments
We thank T.Harada, K. Koyama, S. Kuruyonagi, T. Tanaka, D. Wands, J. Yokoyama  and C. Yoo for useful discussions and feedback. 
This work was supported in part by grant-in-aid for scientific research on innovative areas No. 24103006, No. 21111006, No. 22244030, and No. 23540327, JSPS Grant-in-Aid
for Scientific Research (A) No. 21244033 and MEXT Grant-in-Aid for Scientific Research on Innovative Areas No. 24111701 

\appendix
\section{The time integral during the radiation era}
\label{app:rad-time}
\bea\label{t1}
\tilde{I}_1&=&\frac{1}{4ky^3v^3}\left\{-\cos(x)\sum_{n=1}^{4}\alpha_n\rm{Si}(\beta_nx)
  +\sin(x)\sum_{n=1}^{4}(-1)^{n+1}\alpha_n\rm{ci}(\beta_nx)\right\}\nonumber\\
&&+\gamma_1\sin(x)+\gamma_2\sin(vx)\sin(yx)+\gamma_3\sin(vx)\cos(yx)+
\gamma_4\cos(vx)\sin(yx)+\gamma_5\cos(vx)\cos(yx)\nonumber\\
&& \eea and the second integral is given by: \bea\label{t2}
\tilde{I}_2&=&-\frac{\alpha}{2kv^3y^3}\left\{\cos(x)\left[-\rm{Si}(\beta_1x)+\rm{Si}(\beta_2x)+\rm{Si}(\beta_3x)-\rm{Si}(\beta_4x)
  \right]\right.\nonumber\\
&&\left.+\sin(x)\left[\rm{ci}(\beta_1x)+\rm{ci}(\beta_2x)-\rm{ci}(\beta_3x)-\rm{ci}(\beta_4x)\right]\right\}\nonumber\\
&&+\gamma_{21}\sin(x)+\gamma_{22}\sin(x(v+y))+
\gamma_{23}\sin(x(v-y))+\gamma_{24}\cos(x(v-y))+\gamma_{25}\cos(x(v+y))\nonumber\\
&& \eea 
the coefficients in these two integrals are given in the tables below.  The $\alpha_n$ coefficients in
the first integral satisfy
$\sum_{n=1}^{4}(-1)^n\alpha_n=0$  and  the $\alpha$
coefficient in the second integral is given by $\alpha=(v^2-1+y^2)^2$.

The $\rm{Si}$ and $\rm{ci}$ terms are the sine and cosine integrals respectively \cite{abr}, defined as:

\bea
\rm{Si}(x)&=&\int_0^x\frac{\sin(t)}{t}dt\nonumber\\
\rm{ci}(x)&=&\int_0^x\frac{\cos(t)}{t}dt~.
\eea

\begin{table}[h]
\begin{center}
\begin{tabular}{|l|c|}
\hline
$\beta_1$&$1+v+y$\\
$\beta_2$&$-1+v+y$\\
$\beta_3$&$1+v-y$\\
$\beta_4$&$-1+v-y$\\
\hline
\end{tabular}
\begin{tabular}{|c|c|c|}
\hline
Coefficient&Symbol&Expression\\
\hline
$\sin(x)$&$\gamma_1$&$\frac{1}{kv^2y^2}(v^2-3y^2+1)$\\
$\sin(vx)\sin(yx)$&$\gamma_2$&$\frac{1}{kx^3y^3v^3}(2-x^2-x^2y^2+3x^2v^2)$\\
$\sin(vx)\cos(yx)$&$\gamma_3$&$-\frac{2}{ky^2x^2v^3}$\\
$\cos(vx)\sin(yx)$&$\gamma_4$&$-\frac{2}{kx^2y^3v^2}$\\
$\cos(vx)\cos(yx)$&$\gamma_5$&$\frac{2}{kxy^2v^2}$\\
\hline
$\sin(x)$&$\gamma_{21}$&$\frac{2}{kv^2y^2}(1-v^2-y^2)$\\
$\sin(x(v-y))$&$\gamma_{22}$&$\frac{2}{kx^2v^3y^3}(v-y)$\\
$\sin(x(v+y))$&$\gamma_{23}$&$-\frac{2}{kx^2v^3y^3}(v+y)$\\
$\cos(x(v-y))$&$\gamma_{24}$&$-\frac{1}{kx^3y^3v^3}(-2+x^2-x^2y^2-x^2v^2-2x^2vy)$\\
$\cos(x(v+y))$&$\gamma_{25}$&$-\frac{1}{kx^3v^3y^3}(2-x^2+x^2y^2+x^2v^2-2x^2vy)$\\
\hline
\end{tabular}
\end{center}
\caption{Table on the left gives the expressions for the coefficients of 
the arguments of the Cosine and Sine integrals in \eqs{t1}{t2}.
The right table gives the expressions for the $\tau_1$ integral \eq{t1} (tob block)
and for the $\tau_2$ integral \eq{t2} (bottom block).}
\label{beta}
\end{table}

\begin{table}
\centering
 \begin{tabular}{|l|c|c|c|c|c|c|c|}
 \hline
&$1$&$v^4$&$4v^3$&$4v^2$&$3y^4$&$4y^3$&$2y^2v^2$\\
\hline
$\alpha_1$&$-$&$+$&$+$&$+$&$-$&$-$&$-$\\
$\alpha_2$&$+$&$-$&$+$&$-$&$+$&$-$&$+$\\
$\alpha_3$&$+$&$-$&$-$&$-$&$+$&$-$&$+$\\
$\alpha_4$&$-$&$+$&$-$&$+$&$-$&$-$&$-$\\
\hline
\end{tabular}
 \caption{This table gives the expressions of the coefficients of the sine and cosine
integrals in \eq{t1}. Each $\alpha_n$ coefficient has the same parameters
as the others, but the parameters differ in their respective signs. 
The columns to the right of the $\alpha$s give the sign of the parameter defined
in the column header. For example then we can read off $\alpha_1$ as $-1+v^4+4v^3+4v^2-3y^4-4y^3-2y^2v^2$.}
\label{tab:alpha}
\end{table}

\bibliographystyle{JHEPmodplain}
\bibliography{MDGW3}

\end{document}